\documentclass[epj]{svjour}
\usepackage{graphicx}
\usepackage{amsmath}

\newcommand{\be}{\begin{equation}}
\newcommand{\ee}{\end{equation}}
\newcommand{\ben}{\begin{eqnarray}}
\newcommand{\een}{\end{eqnarray}}

\newcommand{\nd}{\noindent}
\begin{document}

\title{MaxEnt and dynamical information}

\author{
A. Hernando\inst{1}\thanks{alberto.hernando@irsamc.ups-tlse.fr},
A. Plastino\inst{2,3}\thanks{plastino@fisica.unlp.edu.ar} \and 
A. R. Plastino\inst{2,4}\thanks{plastino@sinectis.com.ar}}

\institute{Laboratoire Collisions, Agr\'egats, R\'eactivit\'e,
IRSAMC, Universit\'e Paul Sabatier 118 Route de Narbonne 31062 -
Toulouse CEDEX 09, France \and Instituto Carlos I de F\'isica
Te\'orica y Computacional and Departamento de F\'isica At\'omica,
Molecular y Nuclear Universidad de Granada, Granada, Spain \and
National University La Plata, Physics Institute (IFLP-CCT-CONICET)
C.C. 727, 1900 La Plata, Argentina \and CREG-National University
La Plata-CONICET C.C. 727, 1900 La Plata, Argentina}

\abstract{The MaxEnt solutions are shown to display a variety of
behaviors (beyond the traditional and customary exponential one)
if adequate dynamical information is inserted into the concomitant
entropic-variational principle. In particular, we show both
theoretically and numerically that power laws and power laws with
exponential cut-offs emerge as equilibrium densities in
proportional and other dynamics.}

% 89.70.Cf  ------- Entropy and other measures of information
% 05.90.+m  ------- Other topics in statistical physics, thermodynamics, and nonlinear dynamical systems
% 05.45.Df  ------- Fractals
% 02.50.-r  ------- Probability theory, stochastic processes, and statistics
%89.75.-k Complex systems
%89.75.Da Systems obeying scaling laws
%89.75.Fb Structures and organization in complex systems
%89.75.Hc Networks and genealogical trees
%89.75.Kd Patterns


\date{\today}

\titlerunning{MaxEnt and dynamical information}
\authorrunning{A. Hernando et al.}

\maketitle

\section{Introduction}
\nd The principle of maximum entropy is a fundamental idea of
contemporary science. It states that, subject to known
constraints, the probability distribution which best represents
the current state of knowledge is the one with largest entropy
\cite{jaynes,katz}. In other words, let some testable information
about a probability distribution function be given and consider
the set of all trial probability distributions that encode this
information. The probability distribution that maximizes the
information entropy should be regarded as the optimal probability
distribution with respect to the a priori available information.
In most practical cases, the testable information is given by a
set of conserved quantities (average values of some moment
functions), associated with the probability distribution in
question. This is the way the maximum entropy principle is most
often used in statistical thermodynamics. Another possibility is
to prescribe some symmetries of the probability distribution. An
equivalence between the conserved quantities and corresponding
symmetry groups implies the same level of equivalence for both
these two ways of specifying the testable information in the
maximum entropy method. The maximum entropy principle is also
needed to guarantee the uniqueness and consistency of probability
assignments obtained by different methods, statistical mechanics
and logical inference in particular. The maximum entropy principle
makes explicit our freedom in using different forms of prior
information. As a special case, a uniform prior probability
density (Laplace's principle of indifference) may be adopted.
Thus, the maximum entropy principle is not just an alternative to
the methods of inference of classical statistics, but it is an
important conceptual generalization of those methods \cite{katz}.

\nd In this communication we reveal how to
accommodate dynamical information into the principle via a special
treatment of the equations of motion that considers proportional
and larger-than-proportional dynamics due to their importance in
the study of complex systems. Some rather surprising results ensue
(power-laws and power-laws with exponential cutoffs \cite{power})
that illustrate the power of the approach.

\nd We demonstrate that {\it taking into account  dynamical
information within MaxEnt} involves adding to the pertinent
Hamiltonian new terms and that these {\it resemble the so-called
``information cost'' lucidly introduced by the authors of Ref.
\cite{X1}}. In this way we explicitly reconcile two apparent
different viewpoints, i.e., that of the growth models of Simon
\cite{X2} and the information-treatment of Mandelbrot \cite{X3},
showing that the equilibrium density of a growth process is the
one that maximizes the entropy associated to the enlarged
Hamiltonian introduced here.

\nd Our presentation is organized as follows: in Section 2 we
present the basics of the problem; in Section 3 we describe the
theoretical approach finding the equilibrium densities by means of
MaxEnt according to the dynamical equation that governs the system
at habd; in Section 4 we confirm our findings by means of
numerical experiments with random walkers, and we close drawing
some conclusions in Section 5.

\section{Preliminary matters}
\nd      Let us define
\begin{description}
  \item[i)]   $N$ as the total number of elements/members of a discrete set,
  \item[ii)]  $n_c$ as the total number of special subsets into which the $N-$elements can be grouped,
  \item[iii)] $x_i$ as the number of members of the $i$-th subset,
  \item[iv)]  $n(x)$ as the total number of subsets with exactly $x$ members.
\end{description}
\nd Considering that $n(x)$ is a discrete distribution, the
conservation of both $N$ and $n_c$ guarantees
\begin{description}
  \item[i)]  $\displaystyle \sum_{x=1}^\infty n(x)=n_c$, and
  \item[ii)] $\displaystyle \sum_{x=1}^\infty x n(x)=N$.
\end{description}
\nd Let us now consider the continuous limit of the distribution
$n(x)/n_c\rightarrow p(x)dx$.

\nd     Our goal here is that of finding out, via MaxEnt,  the
explicit form of $p(x)dx$ from either some simple
expectation-values' constraints or, and this is the novelty, from
dynamical information not of that kind.

\section{Theoretical approach}

\subsection{Brownian motion: the ideal gas}

\nd We first consider, as a control case, the dynamics behind the
evolution of subsets of sizes $x$ via the  linear equation:
\begin{equation}
\dot{x}=k
\end{equation}
where $k$ involves a Wiener process, i.e.,  $\langle
k_i(t)k_j(t')\rangle=K\delta_{ij}\delta(t-t')$ being $K$ the
variance of $k$. We are in the presence of the well-know Brownian
motion, which obeys the diffusion equation
\begin{equation}
\partial_t p(x,t) = D\partial_x^2 p(x,t),
\end{equation}
with $D$ the diffusion coefficient and $p(x,t)$ the
non-equi\-li\-brium density at instant $t$. Starting at $t=0$ with
a Dirac-delta distribution $p(x,0)=\delta(x-x')$, the solution to
this equation is  a gaussian distribution of the form (we set for
simplicity $x'=0$)
\begin{equation}
p(x,t) = \frac{1}{\sqrt{4\pi Dt}}e^{-\frac{(x)^2}{4Dt}}.
\end{equation}

\nd The Shannon entropy measure is defined (up to a constant) as
\begin{equation}
S = -\int_\Omega dx p(x)\log p(x),
\end{equation}
where $\Omega$ is a ``volume'' in $x-$space defined by a lower and
an upper limit of allowed sizes $x_0$ and $x_M$, respectively
($x_0\leq x\leq x_M$, thusly $\Omega=x_M-x_0$,).
As our first trial we use just MaxEnt with
a normalization constraint
\begin{equation}
\delta\left[ S - \mu\int_\Omega dx p(x) \right]=0,
\end{equation}
where $\mu$ is the associated Lagrange multiplier. The density
that extremizes this quantity is the constant one $p(x)dx =
Z^{-1}\,dx$ with $Z=\Omega$ the normalization constant (and also
by definition the partition function \cite{katz}), representing
the classical ideal gas ---an ensemble of non-interacting
particles at constant density and gaussian distribution of
velocities. As a second trial we add a constraint on the first
moment (the mean value of the sizes) $\langle x \rangle = N/n_c$,
\begin{equation}
\delta\left[ S - \mu\int_\Omega dx p(x) -\lambda\int_\Omega dx
p(x) x \right]=0.
\end{equation}
The concomitant distribution  is the well-know exponential density
\be \label{Z1} p(x)dx=\frac{\exp(-\lambda x)dx}{Z}, \ee  followed
by an ideal gas in the gravitational field. Note that $x_M$ could
diverge here, but $x_0$ has a lower bound. The constants $Z$
and $\Lambda=x_0\lambda$ (defined for convenience) can be obtained
according to
\begin{equation}
\begin{array}{c} \label{firstZ}
\displaystyle Z = x_0 \frac{e^{-\Lambda}}{\Lambda}\\\\
\displaystyle \Lambda^{-1}+1 = \frac{N}{n_c x_0}.
\end{array}
\end{equation}
We depict in Fig. 1 an arbitrary case for $x_0=1$, $N=250000$ and
$n_c=100000$ ($\Lambda=1/24$).

%%%%%%%%%%%%%%%%%%%%%%%%%%%%%%%%%%%%%%%%%
\begin{figure}[t]
\begin{center}
\includegraphics[width=0.45\textwidth]{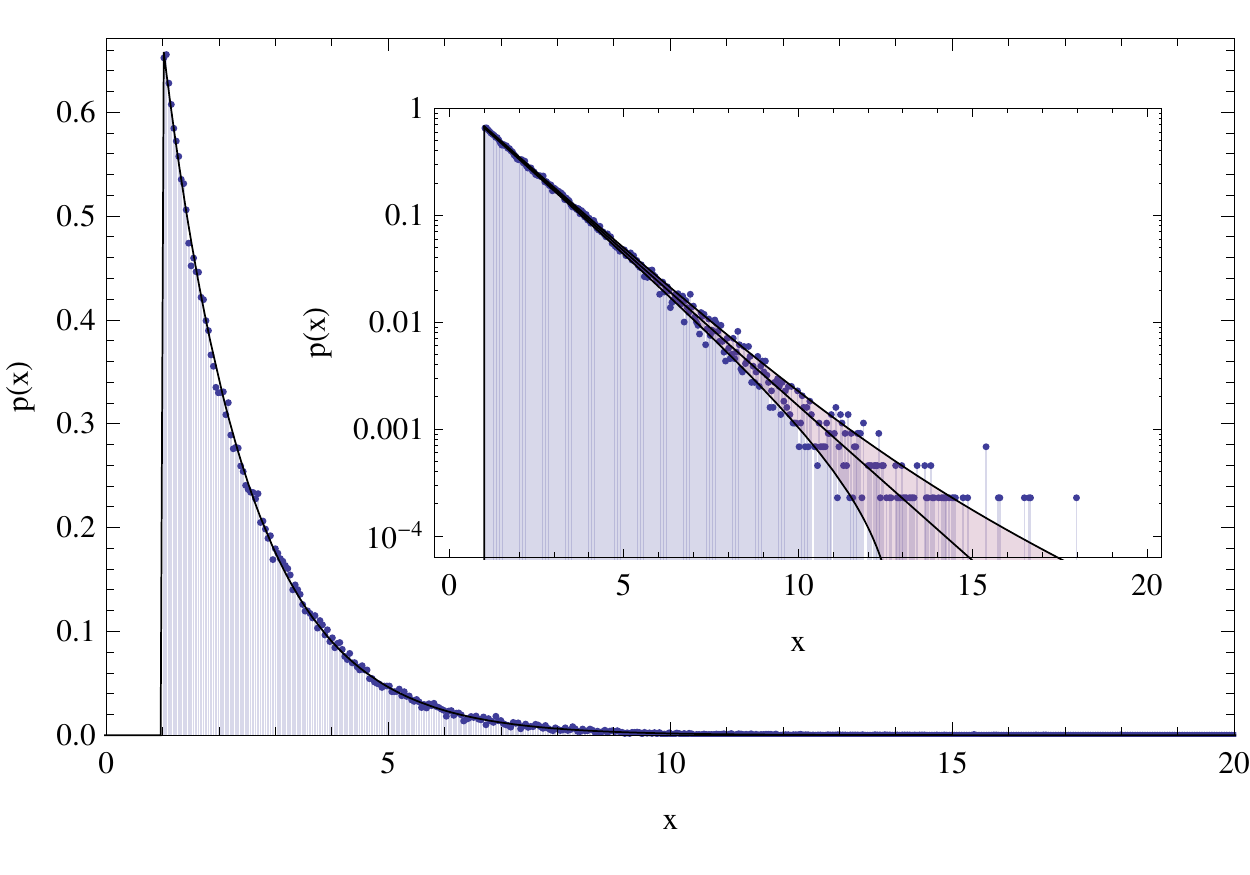}
\caption[]{Equilibrium density for the $q=0$ case (see Eq.
(\ref{plott}) in the text) for $x_0=1$, $N=250000$ and
$n_c=100000$ ($\Lambda=1/24$). Black line: prediction from the
MaxEnt, blue dots: histogram of the asymptotic distribution of
brownian walkers. The shadowed area in the inset around the line
reflects the numerical error of the simulation $O(1/\sqrt{n_c})$.}
\label{fig2}
\end{center}
\end{figure}
%%%%%%%%%%%%%%%%%%%%%%%%%%%%%%%%%%%%%%%%%

 \nd {\it Let us  assume that our constraint is a mean-energy-one}. Then,
 the distribution (\ref{Z1}) can  be associated to a
``Hamiltonian" \cite{katz}
 \be \label{H1}
 H= \Lambda x/x_0,
 \ee
   and is, for instance, the distribution followed by an
ideal gas in the gravitational field.
   Having a Hamiltonian, it is straightforward to introduce a
 temperature here by multiplying it by $\beta=1/T$. The partition
 function defined as $Z=\int dx\,\exp{(-\beta\,H)}$ [Eq. (\ref{firstZ})]
 seemingly remains invariant but with a redefined
 $\Lambda$ that changes in the fashion $\Lambda\rightarrow\beta\Lambda$.

\subsection{Geometric Brownian motion: the scale-free ideal gas}

\nd It is well-known that some social and economic systems display
a scale-free behavior [see, for instance,
\cite{oursnow1,oursnow2,oursnow3}, and references therein]. Thus,
we attempt now  introducing a proportional growth into the
orthodox Jaynes-MaxEnt treatment. We start by considering the
equation,
\begin{equation}
\dot{x}=kx
\end{equation}
where $k$ is again indicative of  a Wiener process, and, of
course, we deal here with the very the definition of geometric
Brownian motion. We now proceed to linearize the dynamic equation
via the new variable $u=\log(x/x_0)$ obtaining thereby
\begin{equation}
\dot{u}=k.
\end{equation}
\nd In analogy with the precedent case, we have the usual Brownian
motion for the variable $u$ obeying the diffusion equation. As for
the $x$ observable one has
\begin{equation}
\partial_t p(x,t) =
D\partial_x\left(x\partial_x\left(xp(x,t)\right)\right),
\end{equation}
 solved (with an initial Dirac-delta $p(x,0)=\delta(x-x')$) via
the log-normal distribution (we set for simplicity $x'=0$)
\begin{equation}
p(x,t) = \frac{1}{\sqrt{4\pi Dt}x}e^{-\frac{\log^2(x)}{4Dt}}.
\end{equation}

\nd The Shannon entropy in the variable $u$ is written as
\begin{equation}
S = -\int_\Omega du p(u)\log p(u),
\end{equation}
with  $u$ instead of $x$. The first constraint  is expressed, as
usual, as
\begin{equation}
\delta\left[ S - \mu\int_\Omega du p(u) \right]=0,
\end{equation}
 that yields a constant density for $u$ as $p(u)du = Z^{-1}du$, with
 $Z=\Omega=\log(x_M/x_0)$. Changing
back to the observable $x=x_0e^u$ we find
\begin{equation}\label{sfig}
p(x)dx = \frac{1}{Z}\frac{dx}{x},
\end{equation}
which follows the density-behavior  of the scale-free ideal gas
(SFIG), as found before by means of  Fisher's information in
references \cite{oursnow1,oursnow2,oursnow3}.

\nd We remind the reader here of Benford's law (BL) \cite{benford,benford2},
also called the first-digit law. As shown in Ref. \cite{benford2}
by means of a bright heuristic derivation, the distribution that
originates BL for first digits has the form of Eq. \ref{sfig}.
Its occurrence is typical of low self-correlated data with no characteristic size,
and thus agrees with the SFIG-definition of
a non-interacting system with scale invariance.

%%%%%%%%%%%%%%%%%%%%%%%%%%%%%%%%%%%%%%%%%
\begin{figure}[t]
\begin{center}
\includegraphics[width=0.45\textwidth]{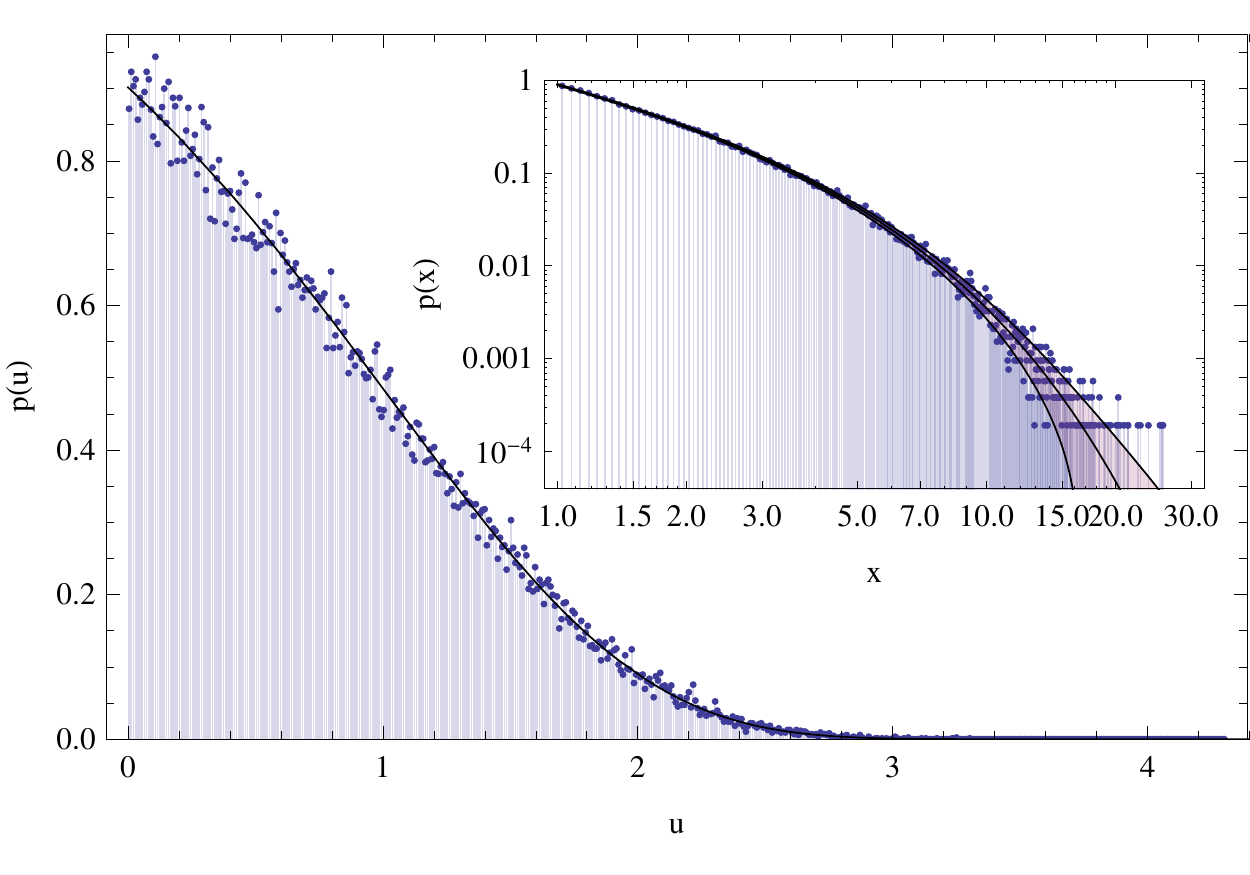}
\caption[]{Same as Fig. 1 for the $q=1$ case of Eq.
(\ref{plott})}. \label{fig3}
\end{center}
\end{figure}
%%%%%%%%%%%%%%%%%%%%%%%%%%%%%%%%%%%%%%%%%

\nd The second constraint is now expressed in terms of $u$ via
$\langle x \rangle = x_0 \langle e^u \rangle = N/n_c$, and the
Jaynes-MaxEnt extremization problem becomes
\begin{equation}
\delta\left[ S - \mu\int_\Omega du p(u) - \Lambda\int_\Omega du
p(u) e^u \right]=0,
\end{equation}
where we have used the definition $\Lambda=x_0\lambda$. We obtain
 the density $p(u)du = Z^{-1}\exp(-\Lambda e^u)du$. Changing back to
the observable $x$ we arrive at

\begin{equation} \label{Z2}
p(x)dx = \frac{1}{Z}\frac{\exp\left(-\Lambda x/x_0\right)}{x}dx.
\end{equation}
The pertinent constants are obtained from the constraints in the
fashion
\begin{equation}\label{z1}
\begin{array}{c}
\displaystyle Z = \Gamma(0,\Lambda)\\
\displaystyle \frac{e^{-\Lambda}}{\Lambda\Gamma(0,\Lambda)} = \frac{N}{n_c x_0},
\end{array}
\end{equation}
where $\Gamma(a,z)$ is the so-called incomplete Gamma function.
This is then the expected equilibrium distribution for an
scale-free system, as those of opinion cluster dynamics in
networks with fixed number of nodes $N$ as well as the number
clusters $n_c$ \cite{oursnow3}. We display in Fig.~2 the case for
the same values of $x_0$, $N$ and $n_c$ as in the preceding
Section (now with $\Lambda=0.360743$).

\nd Assume again that our constrain is a mean-energy-one. We can
associate then to the distribution (\ref{Z2}) the effective
proportional-growth Hamiltonian \cite{katz}

\be \label{H2} H= \Lambda x/x_0 + \ln(x/x_0), \ee   where the new
term $\ln(x/x_0)$ is the dynamical counterpart of the
information-cost of Ref. \cite{X1}.

\subsection{Q-metric Brownian motion: the generalization to
hyper-exponential growth}

\nd We now relax the condition of proportional growth and appeal
to the more general expression
\begin{equation} \label{plott}
\dot{x}=kx^q,
\end{equation}
where $q$ is a dimension-less parameter. It is easy to see that
the two former examples are particular cases corresponding to
$q=0$ (Brownian motion) and $q=1$ (geometric Brownian motion). We
call this new generalization of the dynamics the q-metric Brownian
motion and  proceed to a linearization of the dynamic equation by
introduction of  the  variable
\begin{equation}
u=\log_q(x/x_0),
\end{equation}
where $\log_q(z)$ is the q-logarithm of  Tsallis' statistics
\cite{tsallis}. One finds
\begin{equation}
\dot{u}=k.
\end{equation}
As before, this equation describes the Brownian motion in $u$, and
thus a diffusion equation for $x$ of the form
\begin{equation}
\partial_t p(x,t) =
D\partial_x\left(x^q\partial_x\left(x^qp(x,t)\right)\right),
\end{equation}
whose solution for an initial Dirac-delta $p(x,0)=\delta(x-x')$ is
the q-log-normal distribution
\begin{equation}
p(x,t) = \frac{1}{\sqrt{4\pi
Dt}x^q}e^{-\frac{(\log_q(x/x_0)-\log_q(x'/x_0))^2}{4Dt}}.
\end{equation}

\nd We keep using Shannon's entropy in the MaxEnt approach:
\begin{equation}
S = -\int_\Omega du p(u)\log p(u),
\end{equation}
and the first constraint is expressed, as usual, via
\begin{equation}
\delta\left[ S - \mu\int_\Omega du p(u) \right]=0,
\end{equation}
which yields a constant density for $u$ as $p(u)du = Z^{-1}du$,
with $Z=\Omega=\log_q(x_M/x_0)$. Now,
changing to the observable $x=x_0\exp_q(u)$ we obtain
\begin{equation}
p(x)dx = \frac{1}{Z}\frac{dx}{x^q},
\end{equation}
i.e., a power law. We express now the second constraint as
$\langle x \rangle = x_0 \langle \exp_q(u) \rangle = N/n_c$,
writing
\begin{equation}
\delta\left[ S - \mu\int_\Omega du p(u) -\Lambda\int_\Omega du
p(u) \exp_q(u) \right]=0,
\end{equation}
which solution for its extremization is the density
$p(u)du = Z^{-1}\exp(-\Lambda \exp_q(u))du$.
Changing back to the observable $x$ we now get
\begin{equation} \label{Z3}
p(x)dx = \frac{1}{Z}\frac{\exp\left(-\Lambda x/x_0\right)}{x^q}dx.
\end{equation}
The associated  constants are then obtained from the constraints
and one sees that
\begin{equation}\label{zq}
\begin{array}{c}
\displaystyle Z = (\Lambda/x_0)^{q-1} \Gamma(1-q,\Lambda)\\\\
\displaystyle \frac{\Gamma(2-q,\Lambda)}{\Lambda\Gamma(1-q,\Lambda)}= \frac{N}{n_c x_0}.
\end{array}
\end{equation}
We depict in Fig.~3
the distributions for the same parameters as  Figs. 1 and 2 for
$q=1.5$ and $q=2$, obtaining $\Lambda=0.223742$ and
$\Lambda=0.103808$, respectively.
%%%%%%%%%%%%%%%%%%%%%%%%%%%%%%%%%%%%%%%%%
\begin{figure}[t]
\begin{center}
\includegraphics[width=0.45\textwidth]{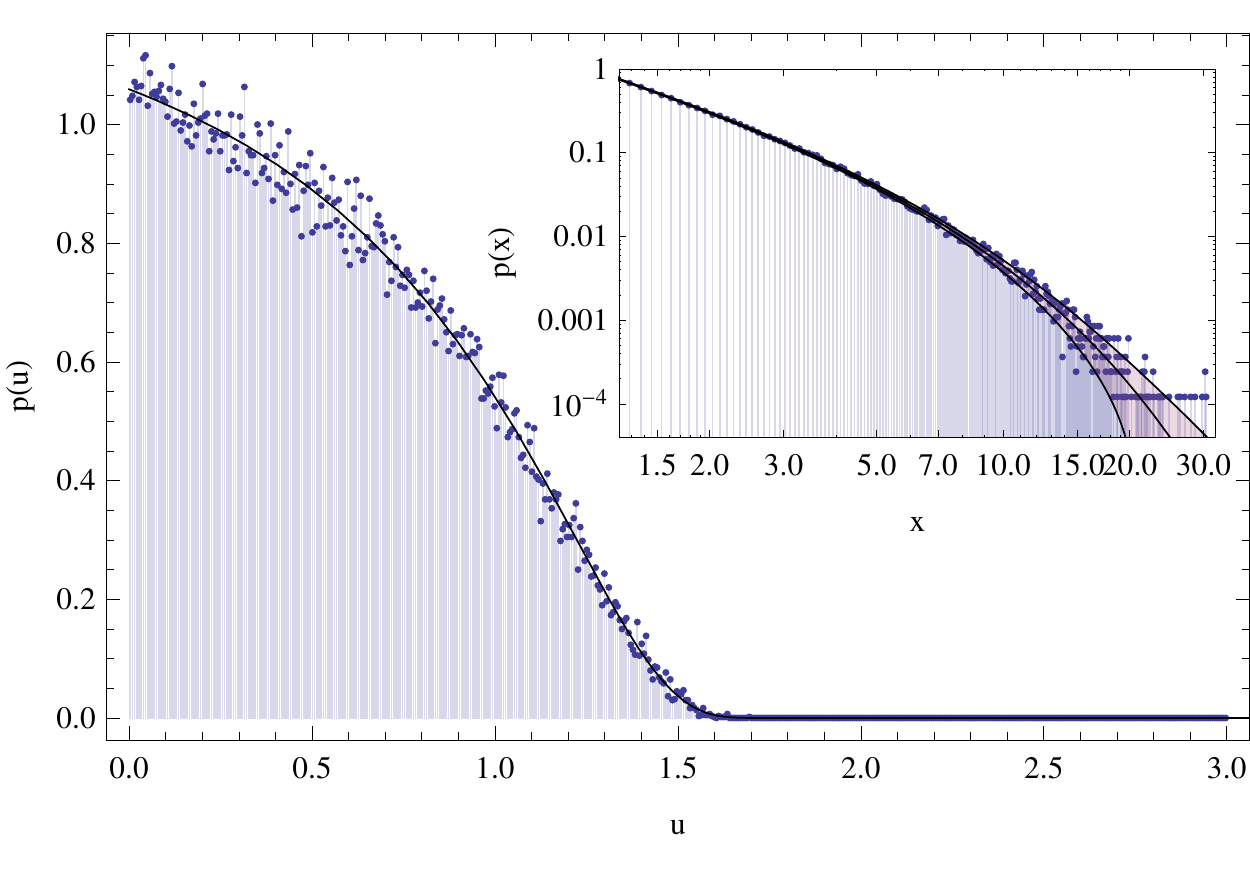}
\includegraphics[width=0.45\textwidth]{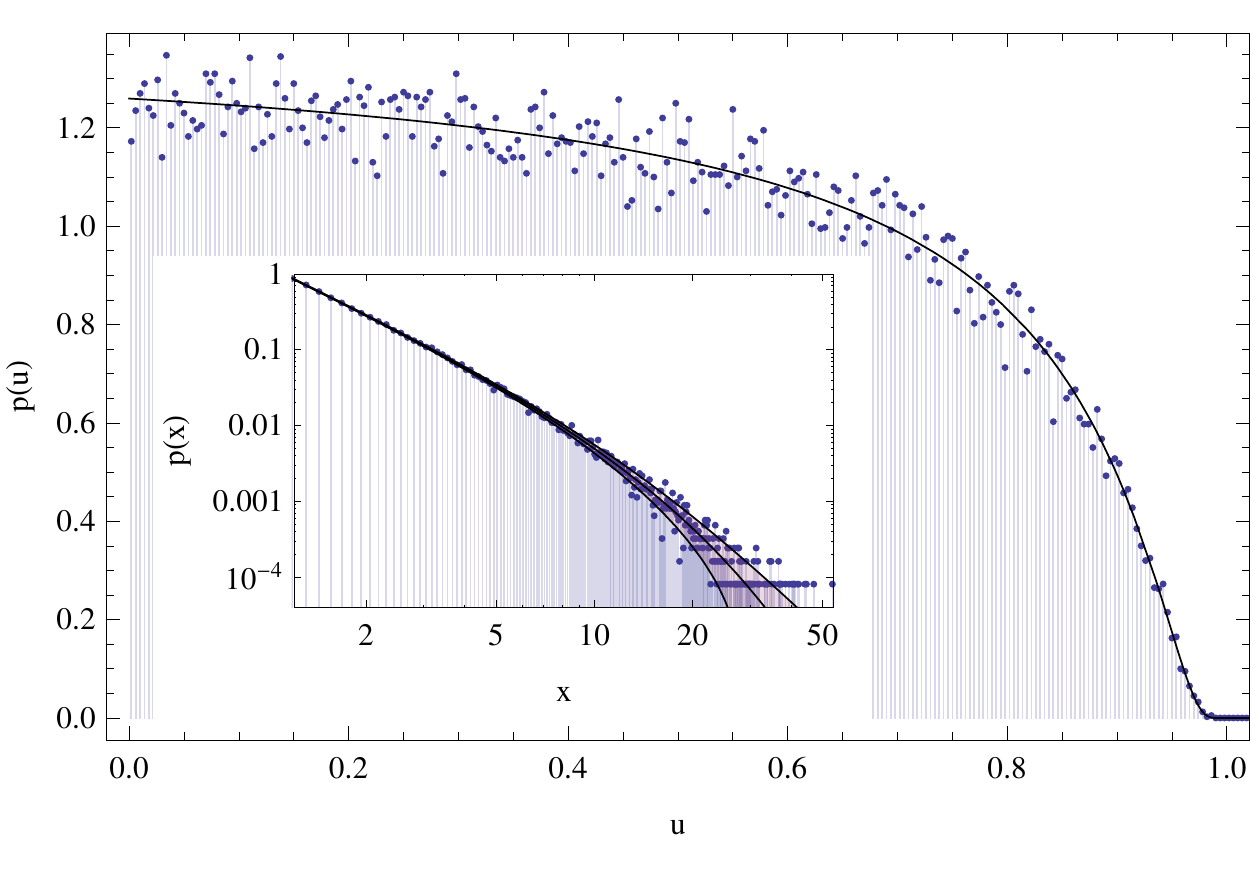}
\caption[]{Same as Fig. 1 for the $q=1.5$ and $q=2$ cases.}
\label{fig4}
\end{center}
\end{figure}
%%%%%%%%%%%%%%%%%%%%%%%%%%%%%%%%%%%%%%%%%

\nd  Assuming once more time that our constrain is a
mean-energy-one, the effective Hamiltonian here reads \cite{katz}
\be \label{H3} H= \Lambda x/x_0\, + q \ln(x/x_0).  \ee Again, the
new term resembles the information-cost of Ref. \cite{X1}. One can
introduce once again an inverse ``temperature" $\beta$ multiplying
the Hamiltonian. The partition function $Z=\int dx\,\exp{(-\beta\,H)}$
turns out to be expressed in terms of the incomplete Gamma
function \cite{abra}
\be \label{beta}
Z=(\beta\Lambda/x_0)^{\beta q-1}\Gamma(1-\beta q,\beta\Lambda).
\ee

\nd Comparing with Eq. (\ref{zq}) the partition function $Z$
remains invariant save for a redefinition $q\rightarrow\beta q$
and $\Lambda\rightarrow\beta\Lambda$. Note also that comparing
with Eq. (\ref{z1}) we recover the proportional growth partition
function at that special temperature for which $\beta=1/q$. Thus,
by increasing $T$ from zero to $q$ we can cancel out a dynamical
behavior via ``heating". Interestingly enough, in the limit
$\beta\rightarrow\infty$ ($T\rightarrow0$) the equilibrium density
distribution is $p(x)dx=\delta(x-x_0)dx$, i.e., all elements
become placed at the same $x_0$. The absolute entropy vanishes, as
it should (third law of thermodynamics). Actual attainment of the
$T=0-$situation would entail a weird configuration indeed, since
it seems impossible that $x_0$ could accommodate all elements
simultaneously.

%%%%%%%%%%%%%%%%%%%%%%%%%%%%%%%%%%%%%%%%%
\begin{figure}[t]
\begin{center}
\includegraphics[width=0.45\textwidth]{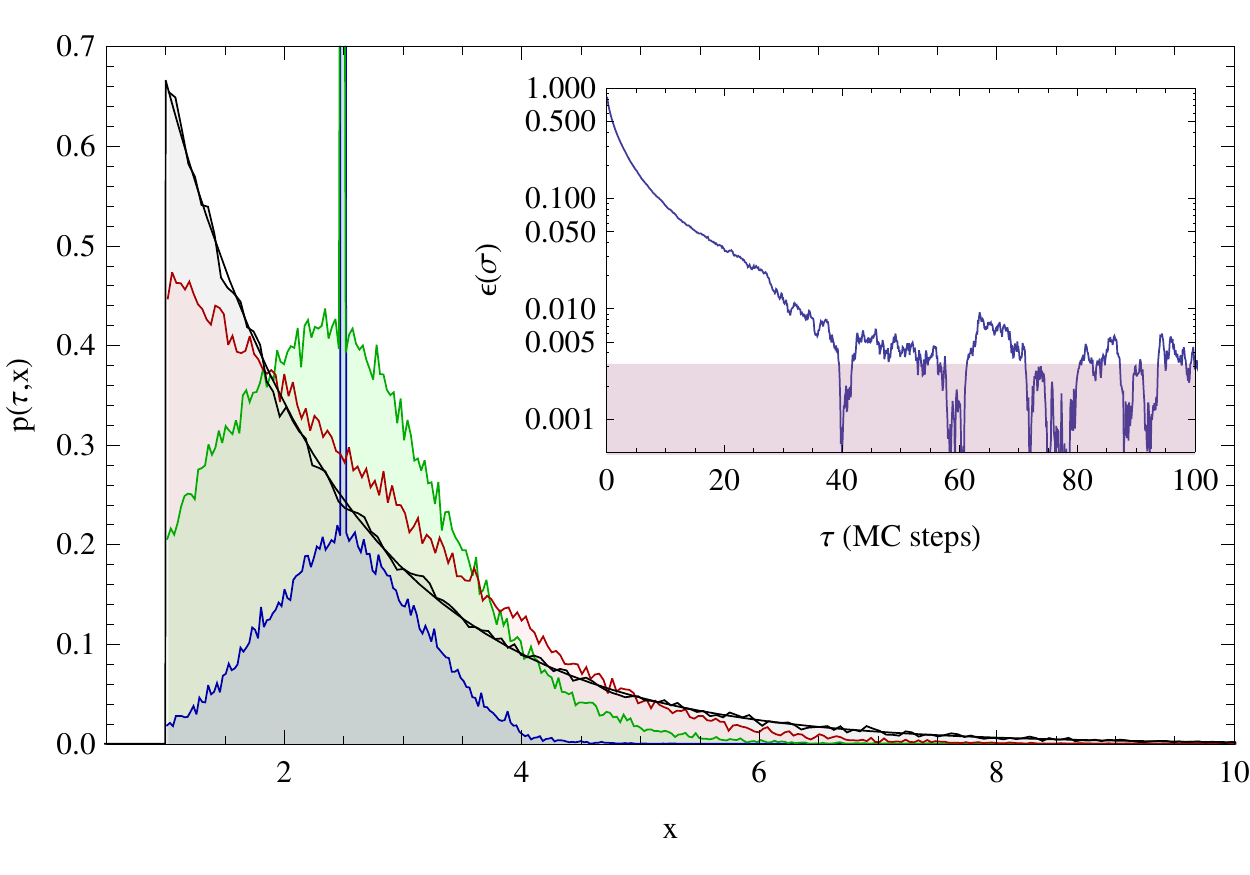}
\caption[]{Diffusion of the initial delta distribution peaked at
$x'=\langle x \rangle = 2.5$, at $\tau=$0.2 (blue), 1.5 (green)
and 4 (red) MC steps, with the asymptotic distribution (black)
compared with the MaxEnt prediction (black smooth line). Inset:
convergence of the standard deviation for the $q=0$ case (see
text). The shadowed area here reflects the numerical error
$1/\sqrt{n_c}$.} \label{fig5}
\end{center}
\end{figure}
%%%%%%%%%%%%%%%%%%%%%%%%%%%%%%%%%%%%%%%%%

\section{Numerical experiments and results}

\subsection{Brownian motion}

\nd We have proceeded to confirm our theoretical findings by means
of numerical experiments, simulating the dynamics of random
walkers following the dynamical equations proposed here. As a
control case, we start with the linear $q=0$, largely used in
physics in molecular dynamics, statistical mechanics and so on.
Our algorithm works as follows:

\begin{description}

\item[i)] We firstly fix the minimum $x_0$, the number of walkers
$n_c$ and the mean value $N/n_c$. We generate a vector
$\mathbf{x}$ with $n_c$ elements as $x_i=N/n_c$, $\forall i$,
which represents the walkers.

\item[iii)] We randomly select the $i$-th walker and generate a
new position by  discretization of the dynamical equation as $x_i
= x_i + k d\tau$, where $k$ is a gaussian random number with
variance $K$ and zero mean, and $d\tau$ is an arbitrary small
`time' interval.

\item[iv)] We correct the mean value in a way compatible with the
dynamical equation, i.e., linearly. A general approach is via the
change $\mathbf{x}' = \mathbf{x} + \Delta$, where
$\Delta=N/n_c-\langle x \rangle$. It is worth  mentioning that the
computational time is reduced by  randomly  choosing a second
$j$-th walker and make it evolve with $x_j = x_j - k d\tau$ using
the same value of $k$ as above. We finally accept the changes if,
and only if, $\min(\mathbf{x})\geq x_0$.

\item[v)] We repeat iii) and iv) iteratively until achieving
convergence in the distribution of $\mathbf{x}$.

\end{description}

\nd     We display in Fig.~5 the diffusion of $n_c=100000$ walkers
initially peaked at $x=2.5$ ($N=250000$), for different simulation
times, at 0.2, 1.5, and 4 Monte-Carlo steps, defining each MC step
as $n_c$ iterations using $K=1$ and $d\tau=10$. We compare them
with the asymptotic equilibrium distribution (also shown in
Fig.~1) and depict the convergence of the relative difference of
the standard deviation to what is  predicted by our  MaxEnt
treatment, $\varepsilon(\sigma) =
|\sigma_{\mathrm{MaxEnt}}-\sigma(\tau)|$. As expected, after some
steps we finally reproduce the theoretical
 MaxEnt distribution. We remark on the importance of correcting the
positions of the walkers in \emph{linear} fashion, respecting the
dynamical equation and guaranteeing maintenance of  the operative
constraint at the main value $\langle x \rangle$.

%%%%%%%%%%%%%%%%%%%%%%%%%%%%%%%%%%%%%%%%%
\begin{figure}[t]
\begin{center}
\includegraphics[width=0.5\textwidth]{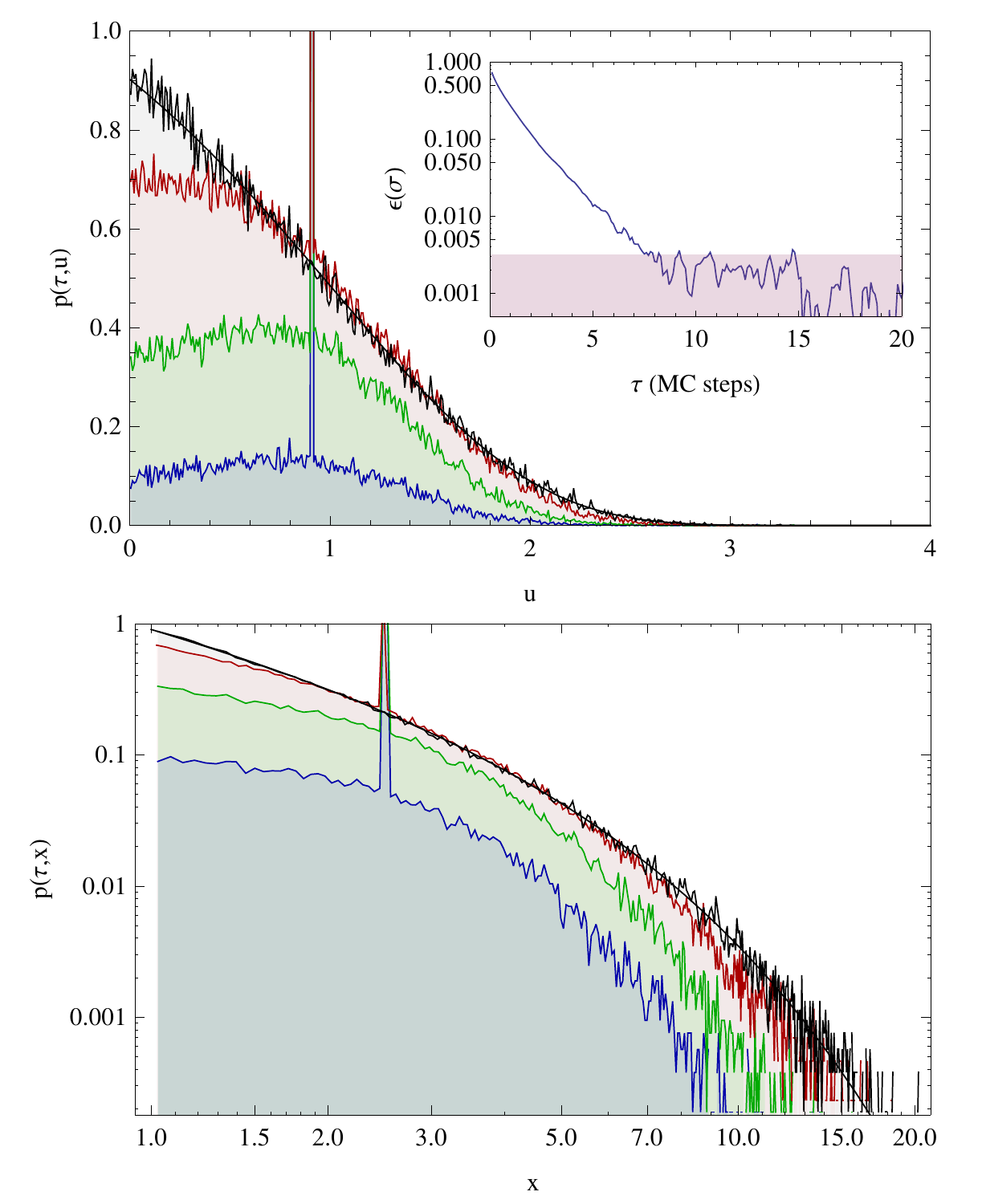}
\caption[]{Same as Fig. 5 for the $q=1$ case, at $\tau=$0.2
(blue), 0.8 (green) and 2.5 (red) MC steps.} \label{fig6}
\end{center}
\end{figure}
%%%%%%%%%%%%%%%%%%%%%%%%%%%%%%%%%%%%%%%%%

\subsection{Geometric brownian motion}

We have proceeded for $q=1$ in a similar way as in the precedent
case, with the difference that now there are two equivalent
descriptions for the dynamics involved. The algorithm for the
first of them is as follows

\begin{description}

\item[i)] We fix the values of $x_0$, $n_c$ and $N/n_c$. We again
generate the walkers as a vector $\mathbf{x}$ with $n_c$ elements
as $x_i=N/n_c$, $\forall i$.

\item[iii)] We randomly select the $i$-th walker and generate a
new position as $x_i = x_i + k x_i d\tau$.

\item[iv)] We now correct the mean value in a way compatible with
the dynamical equation, i.e., proportionally. We change
$\mathbf{x}' = \mathbf{x}\times\Delta$, where now
$\Delta=(N/n_c)/\langle x \rangle$. We accept the changes if, and
only if $\min(\mathbf{x})\geq x_0$.

\item[v)] We repeat iii) and iv) iteratively until encountering
convergence in the distribution of $\mathbf{x}$.

\end{description}

\nd   This algorithm solves explicitly the equation of motion in
$x$, which requires a specially small time interval $d\tau$ in
order to reduce the error in the discretization of the time
derivative. The convergence is achieved after a somewhat big
computational effort since $d\tau\ll(\sqrt{K}x_M)^{-1}$. We highly
recommend working with the variable $u=\log(x/x_0)$ to linearize
the equations, and apply afterwards the forthcoming algorithm:

\begin{description}

\item[i)] We fix the values of $x_0$, $n_c$, and $N/n_c$ to
generate the walkers as a vector $\mathbf{u}$ with
$u_i=\log(N/(n_c x_0))$, $\forall i$.

\item[ii)] We randomly select the $i$-th walker and generate a new
position as $u_i = u_i + k d\tau$.

\item[iii)] We now correct the mean value as $\mathbf{u}' =
\mathbf{u}+\Delta$, where now $\Delta=\log[(N/n_c)/\langle
x_0\exp(u) \rangle]$ ---note that we use now the mean value of the
exponential. We accept the changes if, and only if
$\min(\mathbf{u})\geq 0$.

\item[iv)] We repeat ii) and iii) iteratively until reaching
convergence in the distribution of $\mathbf{u}$.

\end{description}

It is easy to see that both algorithms are equivalents since
$x_i(1+kd\tau)\simeq x_ie^{kd\tau}=e^{u_i+kd\tau}$. We depict in
Fig.~6 the diffusion of the initial delta distribution at 0.2, 0.8
and 2.5 MC steps using $K=1$ and $d\tau=1$ until getting
convergence, with the same values for the parameters as in the
previous instance. The final equilibrium distribution follows that
predicted by MaxEnt, thus demonstrating the validity of our
treatment.

%%%%%%%%%%%%%%%%%%%%%%%%%%%%%%%%%%%%%%%%%
\begin{figure}[ht!]
\begin{center}
\includegraphics[width=0.48\textwidth]{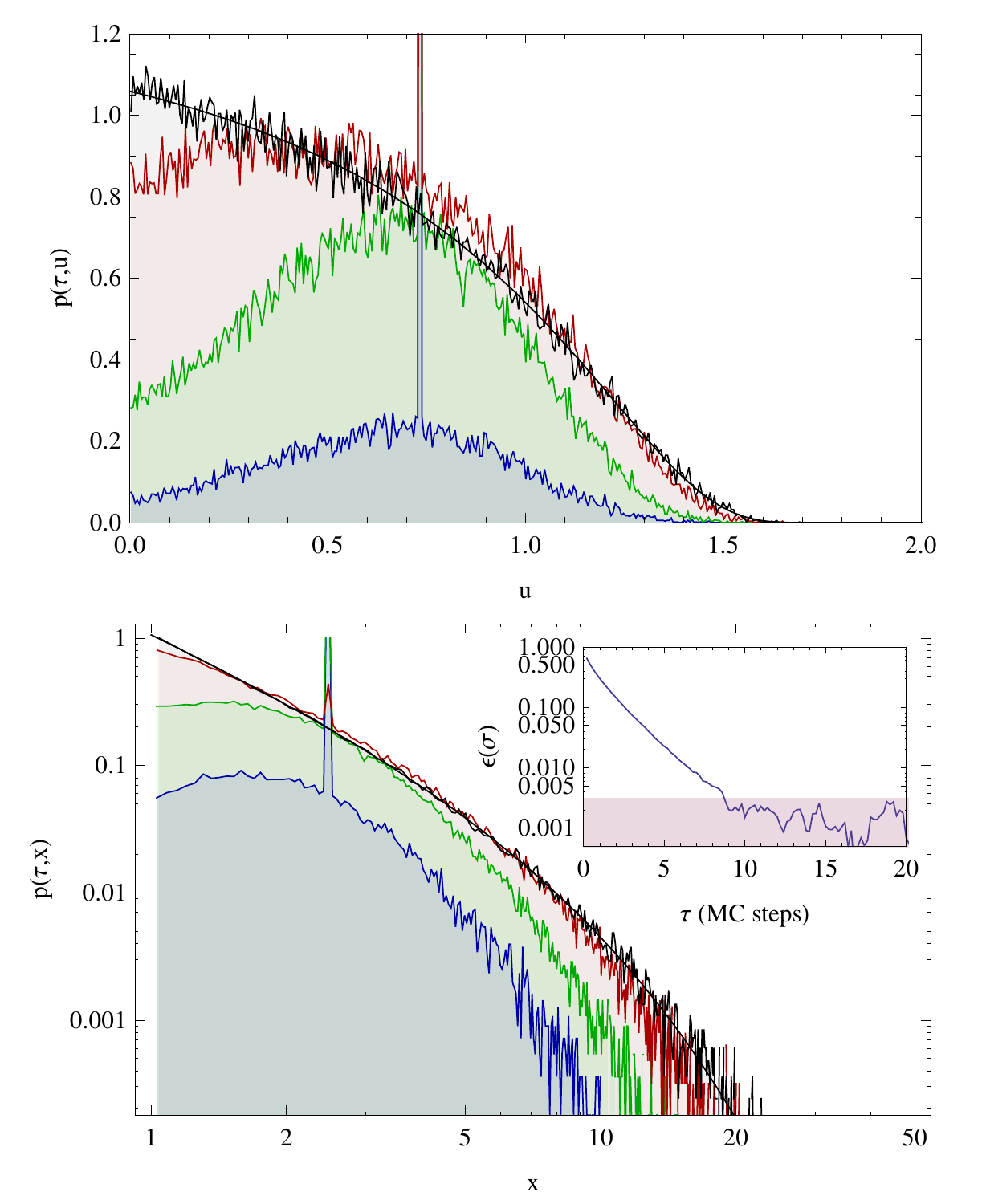}\\
\includegraphics[width=0.48\textwidth]{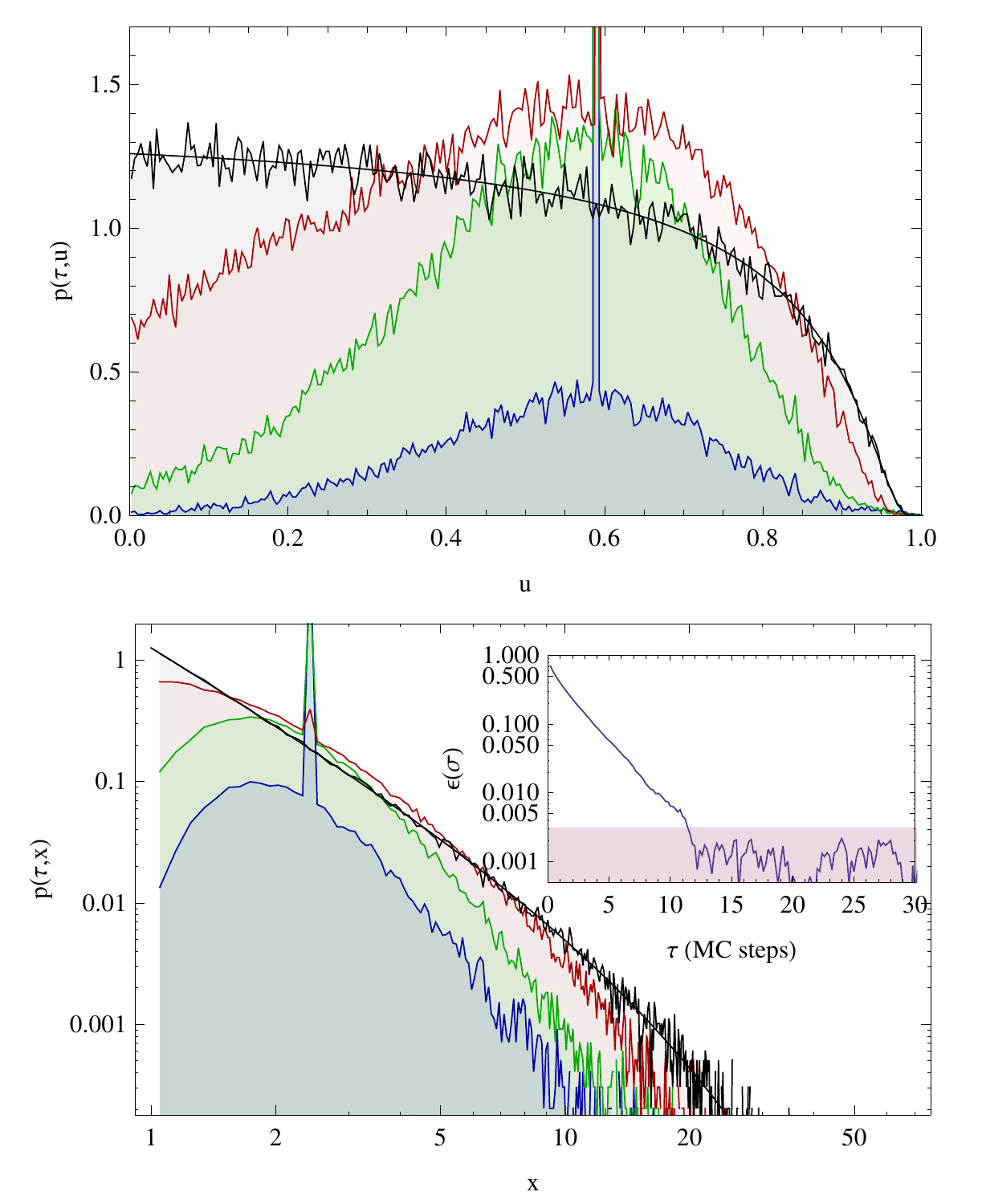}
\caption[]{Top panels: same as Fig. 5 for the $q=1.5$ case at $\tau=$0.1
(blue), 0.5 (green) and 2 (red) MC steps. Bottom panels: $q=2$ case.} \label{fig7}
\end{center}
\end{figure}
%%%%%%%%%%%%%%%%%%%%%%%%%%%%%%%%%%%%%%%%%

\subsection{Q-metric brownian motion}

We finally describe the algorithm   used  for the general case.
 By recourse to  the variable $x$:

\begin{description}

\item[i)] We fix the values of $x_0$, $n_c$, and $N/n_c$ to
generate the vector $\mathbf{x}$ with $x_i=N/n_c$, $\forall i$.

\item[ii)] We randomly select the $i$-th walker and generate a new
position as $x_i = x_i + k x_i^q d\tau$.

\item[iii)] We now correct the mean value as $\mathbf{x}' =
\mathbf{x}+\mathbf{x}^q\Delta$, where now $\Delta=(N/n_c-\langle x
\rangle)/\langle x^q \rangle$, which guarantees obeying the
dynamical equation ---note that we explicitly recover the previous
 cases when $q=0$ and $q=1$. We accept the changes if, and only
if $\min(\mathbf{x})\geq x_0$.

\item[iv)] We repeat ii) and iii) iteratively until convergence in
the distribution of $\mathbf{x}$.

\end{description}

\nd As in the $q=1$ case, this algorithm demands, for a very small
time interval $d\tau$, to reduce the error in the derivative. We
again recommend the use of the linearized variable
$u=\log_q(x/x_0)$ via

\begin{description}

\item[i)] We fix the values of $x_0$, $n_c$, and $N/n_c$ to
generate the vector $\mathbf{u}$ with $u_i=\log_q(N/(n_cx_0))$,
$\forall i$.

\item[ii)] We randomly select the $i$-th walker and generate a new
position as $u_i = u_i + k d\tau$.

\item[iii)] We now correct the mean value as $\mathbf{u}' =
\mathbf{u}+\Delta$. Here $\Delta$ has no analytical expression,
and is obtained from the equation $\langle
x_0\exp_q(u+\Delta)\rangle=N/n_c$ using an iterative Newton
algorithm, reaching convergence in few steps. We accept the
changes if, and only if $\min(\mathbf{u})\geq 0$ and
$\max(\mathbf{u})\leq 1/(q-1)$.

\item[iv)] We repeat ii) and iii) iteratively until obtaining
convergence in the distribution of $\mathbf{u}$.

\end{description}

\nd Using the same parameters as in the previous case, we depict
if Fig.~7 the diffusion process at $0.1$, $0.5$ and $2$
MC steps for the cases $q=1.5$ and $q=2$ respectively, also sowing
the convergence of the standard deviation. Systematically, the
equilibrium densities do follow the predicted distributions found
via the MaxEnt.

\section{Conclusions}

\nd It is commonly believed that Jaynes' MaxEnt (JM), used in
conjunction with Shannon's logarithmic information measure yields,
after the concomitant variational process, only exponential
probability distribution functions (PDF).

\nd   This fact was largely responsible for motivating statistical
mechanics' practitioners to look for other information measures
\cite{tsallis}. We have shown here that great versatility is
gained by MaxEnt if some further, appropriate a priori
``dynamical" knowledge is added to the JM-technique, a way of
proceeding that entirely agrees with Jaynes' philosophy
\cite{jaynes,katz}. Indeed, we see that effective Hamiltonians for
the process at hand are also a result of the MaxEnt technique.

\nd The JM-procedure can in this fashion still keep Shannon's
measure while at the same yielding almost any functional form for
the ensuing variational  PDF, power laws in particular.\\

\nd {\bf ACKNOWLEDGMENT:}  This work was partially supported
by ANR DYNHELIUM (ANR-08-BLAN-0146-01) Toulouse, and the Projects FQM-2445 and FQM-207 of the Junta de Andaluc\'{\i}a.
AP  acknowledges support from the Senior Grant CEI Bio-Tic GENIL-SPR.

\end{document}